\documentstyle[12pt]{article}
\begin{document}
\begin{flushright}
Alberta Thy-01-99\\
January, 1999\\
\end{flushright}
\vskip 1cm

\begin{center}
{\Large \bf  Long-range two-body final-state interactions and direct $CP$ asymmetry in ${B}^{+}\rightarrow {\pi}^{+} {K}^{0}$ decay}\\[10mm]
A. N. Kamal \\[5mm]
{\em Theoretical Physics Institute and Department of Physics,\\ University of Alberta,
Edmonton, Alberta T6G 2J1, Canada.}
\end{center}
\vskip1cm

\begin{abstract}
We present a calculation of the direct $CP$ asymmetry, ${A}_{CP}^{dir}$, for the process ${B}^{+}\rightarrow {\pi}^{+} {K}^{0}$ including the effects of long-range two-body inelastic final-state interactions (FSI). We admit four two-body channels in our calculation: $ B^+ \rightarrow ( {\pi}^{+} {K}^{0}), ( \eta {K}^{+})$, $({D}_{s}^{+} {\bar{D}}^{0})$ and the charge-exchange channel  ${\pi}^{0} {K}^{+}$. The strong scattering is described in terms of Pomeron and Regge exchanges. We find that the direct $CP$ asymmetry is enhanced by a factor of $\sim 3$ as a result of FSI, but remains well short of the claims of $(10 - 20)\%$ in recent literature. A critical assessment of papers claiming large $CP$ asymmetries is also presented. Many-body decay channels could play an important role in direct $CP$ asymmetry; however, their contribution is hard to estimate.
\end{abstract}
PACS categories: 11.30.Er,13.25.Hw.

\newpage

\begin{center}
{\bf I. INTRODUCTION}
\end{center} 

Considerable attention \cite{{flm},{fl1},{nr}} has been paid to $B\rightarrow \pi K$ decays for their potential to determine the $CKM$ angle $\gamma$. The effect of two-body final-state interactions (FSI) has also been studied variously by Refs. \cite{{fl1},{n},{jin},{du}, {falk}}. In calculations \cite{kram} without the inclusion of FSI effects, one finds a direct $CP$ asymmetry, ${{A}_{CP}^{dir}}$, in ${B}^{+}\rightarrow  {\pi}^{+} {K}^{0}$ of the order of 0.5\%. However, Refs. \cite{jin} and \cite{du} calculating the effect of an intermediate  ${\rho} {K}^{*}$ state converting to ${\pi}^{+} {K}^{0}$ through an elementary pion exchange, estimate the direct $CP$ asymmetry arising out of this triangle graph to be $\sim 10\%$. This would imply that such a large $CP$ asymmetry could be accomodated within the Standard Model. Further, it has been argued in Refs. \cite{falk} that in ${B}^{+}\rightarrow  {\pi}^{+} {K}^{0}$ decay a $CP$ asymmetry of the order of $(10-20)\%$ is possible once long-distance (soft) FSI are included.  Hence it is argued  that a $20\%$ $CP$ asymmetry could not be taken unambiguously to signal New Physics. Ref. \cite{neub} has also discussed the role of FSI in the determination of the angle $\gamma$, and concluded that $CP$ asymmetry as large as $15\%$ can be expected in  ${B}^{+}\rightarrow  {\pi}^{+} {K}^{0}$ decay.

	The goal of the present paper is rather limited: it is to investigate the effect of long-range two-body FSI on ${{A}^{dir}_{CP}}$ in the process ${B}^{+}\rightarrow  {\pi}^{+} {K}^{0}$ which in the absence of FSI is a pure penguin-driven process. We admit four two-body channels in the discussion of inelastic final-state interactions: $( {\pi}^{+} {K}^{0}), ( \eta {K}^{+})$,  $({D}_{s}^{+} {\bar{D}}^{0})$ and the charge-exchange (CEX) channel ${\pi}^{0} {K}^{+}$. The decay amplitude is decomposed into the two isospin states, $I=1/2$ and $3/2$, and each corrected for FSI. Our method of including FSI is different from that of \cite{falk}. In contrast to the claims in Ref.  \cite{falk}, we find that ${A}_{CP}^{dir}$ could be raised by a factor of $\approx 3$, but remains well short of (10 - 20)\% range. We spell out the role played by the elastic and inelastic two-body channels on ${{A}^{dir}_{CP}}$, and take a critical look at Refs. \cite{{jin}, {du}, {falk}} thereby tracing also the reasons for the variance between our results and theirs. Following Ref. \cite{falk}, we also neglect the electromagnetic penguin processes.

The layout of the paper is as follows: We define the Hamiltonian and the model parameters in Section II. Weak decay amplitudes for the four two-body channels,  $ B^+ \rightarrow ( {\pi}^{+} {K}^{0}), ( \eta {K}^{+})$, $({D}_{s}^{+} {\bar{D}}^{0})$ and  $({\pi}^{0} {K}^{+})$, are calculated in Section III. The strong $S$-matrix elements, ${S}_{11}, {S}_{12}$, ${S}_{13}$ and ${S}_{14}$, are calculated  in Section IV, using Pomeron and Regge exchanges. $CP$ asymmetry calculations are presented in Section V. The concluding section, Section VI, contains a discussion of the results, and a critique of previous calculations claiming large asymmetries. 
\vskip 1cm

\begin{center}
{\bf II. DEFINITIONS AND PARAMETERS}
\end{center}
	In our analysis we will consider only the the QCD-modified effective Hamiltonian for $b\rightarrow s$ transitions with the top quark integrated out \cite{{buras}, {fl2}, {desh}}

\begin{equation}
H_{eff}={G_F  \over \sqrt{2} } \sum_{q=u,c}{\left\{  V_{qb} V^*_{qs}[ C_1 O^q_1+C_2 O^q_2+
\sum_{i=3}^{6}{C_i O_i}]  \right\} } .
\end{equation}

Here ${C}_{i}$ are the Wilson coefficients defined at scale $\mu\approx {m}_{b}$ and the current bilinears ${O}_{i}$ are defined as follows:

\begin{eqnarray}
& O^q_1   =  (\bar{s}q)_{V-A} (\bar{q}b)_{V-A},   & O^q_2  =  (\bar{s}_{\alpha}q_{\beta})_{V-A} (\bar{q}_{\beta} b_{\alpha})_{V-A}; \nonumber \\
& O_3   =  (\bar{s}b)_{V-A}\sum_{q^{\prime}} (\bar{q}^{\prime}q^{\prime})_{V-A},  & 
O_4  =  (\bar{s}_{\alpha} b_{\beta})_{V-A}\sum_{q^{\prime}} (\bar{q}^{\prime}_{\beta} q^{\prime}_{\alpha})_{V-A},     \nonumber  \\
& O_5  =  (\bar{s}b)_{V-A}\sum_{q^{\prime}} {(\bar{q}^{\prime}q^{\prime})_{V+A}}, & 
O_6  =  (\bar{s}_{\alpha} b_{\beta})_{V-A}\sum_{q^{\prime}}{(\bar{q}^{\prime}_{\beta} q^{\prime}_{\alpha})_{V+A}}.
 \end{eqnarray}

In Eq. (2), $O_1 $ and $O_2 $ are the tree operators, $O_3, ...., O_6 $ are generated by QCD penguin processes,  $(V\pm A )$ represents $\gamma_{\mu} (1\pm \gamma_5)$, and $\alpha$ and $\beta$ are color indices.  $\sum_{q^{\prime}}$ is a sum over the active flavors u,d,s and c quarks.

In the next-to-leading-logarithmic calculation one works with effective Wilson coefficients $C^{eff}_i $, rather than the coefficients that appear in (1). The derivation of these effective coefficients is well known \cite{{buras}, {fl2}, {desh}}. We simply quote their values

\begin{eqnarray}
C^{eff}_1=\bar{C}_1,~~  C^{eff}_2=\bar{C}_2, & ~~ C^{eff}_3=\bar{C}_3 - P_s / N_c,  & C^{eff}_4=\bar{C}_4 + P_s,  \nonumber   \\
C^{eff}_5=\bar{C}_5 - P_s/N_c,~~~~ &  C^{eff}_6=\bar{C}_6 +P_s,
\end{eqnarray}
with\cite{desh}
\begin{eqnarray}
 \bar{C}_1=1.1502, ~\bar{C}_2=-0.3125, ~\bar{C}_3=0.0174, \nonumber \\ \bar{C}_4=-0.0373, ~\bar{C}_5=0.0104, ~ \bar{C}_6=-0.0459,
\end{eqnarray}
and 
\begin{eqnarray}
P_s &= &{\alpha_s (\mu) \over  8\pi} C_1 (\mu) [{10 \over 9}+\frac{2}{3}\ln{\frac{m_q^2}{\mu^2}}-G(m_q, \mu, q^2)],
\end{eqnarray}
where
\begin{equation}
G(m_q, \mu, q^2)= -4\int_{0}^{1}{dx x (1-x) \ln{[1-x (1-x) {q^2 \over  m^2_q}}]}.
\end{equation}
Note that only the current operator ${O}_{1}^{q}$, with q=u,c, is used to generate the penguin contributions; $q^2 $, on the other hand, is the  invariant momentum carried by the gluon in the penguin diagram, and 
$m_q$ the mass of the quark q in the penguin loop.   For  $q^2> 4 m^2_q$,  $G(m_q, \mu, q^2)$ becomes complex giving rise to strong perturbative (short distance) phases through  $P_s $. We use $N_c=3$ throughout.
 
The quark masses and the meson decay constants, we employ are:
\begin{eqnarray}
& m_u= m_d=10 MeV,   m_s=175 MeV   ,   m_c=1.35GeV , m_b=5.0 GeV,    & \nonumber  \\
&  f_\pi=131 MeV  , f_K= 160 MeV, f_{\eta}=130 MeV,   f_{{D}_{s}}= 300 MeV.     &  \end{eqnarray}

As for the $CKM$ angles, Ref. \cite{pdg} quotes $\mid {V}_{ub} \mid/\mid {V}_{cb} \mid = 0.08\pm 0.02$, and Ref. \cite{parodi} quotes Arg $({V}_{ub}^{*}) = {(62\pm12)}^{\circ}$. Refs. \cite{ali} and \cite{desh1} cite a significantly different range for Arg $({V}_{ub}^{*})$. For our calculations, we take,
\begin{equation}
{\mid {V}_{ub} \mid \over \mid {V}_{cb} \mid}=0.1, ~~~~\mbox {and} ~~~ \mbox {Arg}  ({V}_{ub}^{*}) = (50^\circ, 75^\circ, 90^\circ).
\end{equation} 

Since long-range rescattering preserves isospin, we decompose ${B}^{+}\rightarrow {\pi}^{+} {K}^{0}$ amplitude into $I=1/2$ and $I=3/2$ components and then apply Pomeron and Regge exchanges to these amplitudes to compute the FSI-corrected isospin amplitudes. Finally the FSI-corrected isospin amplitudes are recombined to generate the total FSI-corrected decay amplitude which is then used to calculate ${A}_{CP}^{dir}$.
\vskip 1cm
\begin{center}
{\bf  III. DECAY AMPLITUDES}
\end{center}
\begin{center}
{\bf A. Channel ${B}^{+}\rightarrow {\pi}^{+} {K}^{0}$}
\end{center}

The $B\rightarrow \pi K$ decay amplitudes in terms of the isospin amplitudes are given by (see, for example, Ref. \cite{n})
\begin{eqnarray}
{A}_{+0} & = &A({B}^{+}\rightarrow {\pi}^{+} {K}^{0}) = {A}_{3/2} + {A}_{1/2}^{(+)} \nonumber \\
{A}_{0+} & = & \sqrt{2} A({B}^{+}\rightarrow {\pi}^{0} {K}^{+}) = 2{A}_{3/2} - {A}_{1/2}^{(+)} \nonumber \\
{A}_{-+} & = & A({B}^{0}\rightarrow {\pi}^{-} {K}^{+}) = {A}_{3/2} + {A}_{1/2}^{(-)} \nonumber  \\
{A}_{00} & = & \sqrt{2} A({B}^{0}\rightarrow {\pi}^{0} {K}^{0}) = 2{A}_{3/2} - {A}_{1/2}^{(-)},  
\end{eqnarray}

where
\begin{equation}
{A}_{1/2}^{(+)} = {A}_{1/2} + {B}_{1/2}, ~~~ {A}_{1/2}^{(-)} = {A}_{1/2} - {B}_{1/2},
\end{equation}

and
\begin{eqnarray}
{A}_{1/2} & = & \pm\sqrt{2/3}\langle 1/2,\pm1/2| {H}_{\Delta I=1} |1/2,\pm 1/2\rangle \nonumber \\
{B}_{1/2} & = & \sqrt{2/3} \langle1/2,\pm 1/2| {H}_{\Delta I=0}|1/2,\pm 1/2\rangle \nonumber \\
{B}_{3/2} & = & \sqrt{1/3} \langle 3/2,\pm 1/2|{H}_{\Delta I=1} |1/2,\pm 1/2\rangle.
\end{eqnarray}

In Eq. (11) the state vectors are eigenstates of $I$ and ${I}_{3}$. ${H}_{\Delta I=0}$ and ${H}_{\Delta I=1}$ are the isospin-conserving and isospin-changing parts of the Hamiltonian, respectively. Note that the penguin operators, ${O}_{3}$ to ${O}_{6}$, involving ${\sum}_{q}(\bar{q}q)$ are of necessity isospin conserving. The operators ${O}_{1}$ and ${O}_{2}$ generate both ${\Delta I=0}$ and ${\Delta I=1}$ transitions. Note also that there are three independent isospin amplitudes but four decay amplitudes. The constraint is provided by requiring that ${A}_{3/2}$ calculated from ${A}_{+0}$ and ${A}_{0+}$ be the same as that from ${A}_{-+}$ and ${A}_{00}$. This serves to fix the relative signs of the decay amplitudes, as we shall see in the following.

In the factorization approximation the four decay amplitudes of Eq. (9) are given by ( an overall factor of ${{G}_{F}/\sqrt{2}}$  has been suppressed)
\begin{eqnarray}
{A}_{+0} & = & -{f}_{K} ({m}_{B}^{2}-{m}_{\pi}^{2}) {F}_{0}^{B\pi} ({m}_{K}^{2}) \left\{  {\xi}_{u}^{*} 
({a}_{4}^{(u)} +{R}_{d}^{(K)} {a}_{6}^{(u)})\right.  \nonumber \\
& & \left.+{\xi}_{c}^{*}({a}_{4}^{(c)} +{R}_{d}^{(K)}{a}_{6}^{(c)}) \right\},
\nonumber \\
{A}_{0+} & = & {f}_{K}({m}_{B}^{2}-{m}_{\pi}^{2}) {F}_{0}^{B\pi}({m}_{K}^{2}) \left\{{\xi}_{u}^{*} 
\left({a}_{1}+{a}_{4}^{(u)} + {R}_{d}^{(K)} {a}_{6}^{(u)}\right) \right.  \nonumber \\
& & \left.+ {\xi}_{c}^{*}\left({a}_{4}^{(c)} +{R}_{d}^{(K)}{a}_{6}^{(c)}\right)\right\} 
+{f}_{\pi}({m}_{B}^{2}-{m}_{K}^{2}){F}_{0}^{BK}({m}_{\pi}^{2}) {\xi}_{u}^{*} {a}_{2}, \nonumber \\ 
{A}_{-+} & = & {f}_{K}({m}_{B}^{2}-{m}_{\pi}^{2}) {F}_{0}^{B\pi}({m}_{K}^{2}) \left\{{\xi}_{u}^{*} 
\left( {a}_{1} + {a}_{4}^{(u)} + {R}_{u}^{(K)} {a}_{6}^{(u)}\right)\right. \nonumber \\ 
& & \left.+ {\xi}_{c}^{*}\left({a}_{4}^{(c)} +{R}_{u}^{(K)}{a}_{6}^{(c)}\right)\right\}, \nonumber \\
{A}_{00} & = &  -{f}_{K} ({m}_{B}^{2}-{m}_{\pi}^{2}) {F}_{0}^{B\pi}({m}_{K}^{2}) \left\{{{\xi}_{u}}^{*} 
\left({a}_{4}^{(u)} + {R}_{d}^{(K)} {a}_{6}^{(u)}\right)\right.  \nonumber \\ 
& & \left.+{\xi}_{c}^{*}\left({a}_{4}^{(c)} +{R}_{d}^{(K)}{a}_{6}^{(c)}\right)\right\}  +{f}_{\pi}({m}_{B}^{2}-{m}_{K}^{2}){F}_{0}^{BK}({m}_{\pi}^{2}) {\xi}_{u}^{*} {a}_{2},
\end{eqnarray}

where
\begin{eqnarray}
  {a}_{2i}= {C}_{2i}^{eff} +{1 \over {N}_{c}}{C}_{2i-1}^{eff},~~~~~~~~~  \nonumber \\
  {a}_{2i-1}={C}_{2i-1}^{eff}+{1 \over {N}_{c}}{C}_{2i}^{eff},~~~
({N}_{c}=3, i=1,...,6)   \nonumber \\
  {R}_{q}^{(K)}={2{m}_{K}^{2} \over ({m}_{b}-{m}_{q})({m}_{s}+{m}_{q})}, ~~~  {\xi}_{q}={V}_{qb}{{V}_{qs}}^{*},   
\end{eqnarray}
and the superscript on ${a}_{4}^{(q)}$ and ${a}_{6}^{(q)}$ represents the quark flavor in the penguin loop.

The isospin amplitudes derived from the above equations are (note that one has to set ${R}_{u}^{(K)}={R}_{d}^{(K)}$ for isospin symmetry)
\begin{eqnarray}
{A}_{3/2} & = & {1 \over 3} {\xi}_{u}^{*}  \left\{({m}_{B}^{2}-{m}_{\pi}^{2}){f}_{K}{F}_{0}^{B\pi}({m}_{K}^{2}){a}_{1}
+({m}_{B}^{2}-{m}_{K}^{2}){f}_{\pi}{F}_{0}^{BK}({m}_{\pi}^{2}){a}_{2}\right\} \nonumber \\
{A}_{1/2}^{(+)} & = & -{1 \over 3} {\xi}_{u}^{*} \left\{({m}_{B}^{2}-{m}_{\pi}^{2}){f}_{K}{F}_{0}^{B\pi}({m}_{K}^{2})  \left({a}_{1}
+3{a}_{4}^{(u)}+ 3{R}_{d}^{(K)}{a}_{6}^{(u)}\right)\right.  \nonumber \\
 &  & \left.+({m}_{B}^{2}-{m}_{K}^{2}){f}_{\pi}{F}_{0}^{BK}({m}_{\pi}^{2}){a}_{2}\right\} \nonumber \\
 &  & -{\xi}_{c}^{*}({m}_{B}^{2}-{m}_{\pi}^{2}){f}_{K}{F}_{0}^{B\pi}({m}_{K}^{2})  \left({a}_{4}^{(c)} +{R}_{d}^{(K)}{a}_{6}^{(c)}\right) \nonumber \\
{A}_{1/2}^{(-)} & = & {1 \over 3} {\xi}_{u}^{*} \left\{({m}_{B}^{2}-{m}_{\pi}^{2}){f}_{K}{F}_{0}^{B\pi}({m}_{K}^{2}) \left(2{a}_{1}+3{a}_{4}^{(u)}+ 3{R}_{d}^{(K)}{a}_{6}^{(u)}\right)\right.  \nonumber \\
 &  &\left. -({m}_{B}^{2}-{m}_{K}^{2}){f}_{\pi}{F}_{0}^{BK}({m}_{\pi}^{2}){a}_{2}\right\}\nonumber \\
 &  & +{\xi}_{c}^{*}({m}_{B}^{2}-{m}_{\pi}^{2}){f}_{K}{F}_{0}^{B\pi}({m}_{K}^{2})  \left({a}_{4}^{(c)} +{R}_{d}^{(K)}{a}_{6}^{(c)}\right). 
\end{eqnarray}

Notice that ${A}_{3/2}$ gets contribution from the operators ${O}_{1}$ and ${O}_{2}$ only. For the discussion of ${B}^{+}\rightarrow {\pi}^{+} {K}^{0}$ decay only ${A}_{3/2}$ and ${A}_{1/2}^{(+)}$ are relevant. With the chosen parameters of Eq. (7), and the form factors of \cite{bsw}, they are (aside from a factor ${{G}_{F} \over \sqrt{2}}$):
\begin{eqnarray}
{A}_{3/2} & \equiv &  {\xi}_{u}^{*} {x}_{u}^{(3/2)}+  {\xi}_{c}^{*} {x}_{c}^{(3/2)} \nonumber \\
& = & {\xi}_{u}^{*} (0.5529)~~ {GeV}^{3} \nonumber \\
{A}_{1/2}^{(+)} & \equiv  & {\xi}_{u}^{*} {x}_{u}^{(1/2)}+  {\xi}_{c}^{*} {x}_{c}^{(1/2)} \nonumber \\ & =  & {\xi}_{u}^{*} (-0.4637 + 0.04127 i) +  {\xi}_{c}^{*} (0.1034 +0.03595 i)~~{GeV}^{3}.
\end{eqnarray}
Note that  ${x}_{c}^{(3/2)} = 0$. The two isospin amplitudes are unitarized through isospin conserving FSI. Henceforth, the superscript $(+)$ in ${A}_{1/2}^{(+)}$ will be dropped wherever it is possible to do so without ambiguity, since it is the only relevant isospin $1/2$ amplitude.
\vskip 1cm

\begin{center}
{\bf B. Inelastic Channels ${B}^{+}\rightarrow \eta {K}^{+}$, ${B}^{+}\rightarrow {D}_{s}^{+} {\bar{D}}^{0}$ and ${\pi}^{0}{K}^{+}$ }
\end{center}

We specifically consider the effect of three channels, ${B}^{+}\rightarrow \eta {K}^{+}$, ${B}^{+}\rightarrow {D}_{s}^{+} {\bar{D}}^{0}$ and the charge-exchange channel ${B}^{+}\rightarrow{\pi}^{0}{K}^{+}$, on the elastic channel  ${B}^{+}\rightarrow {\pi}^{+} {K}^{0}$. The first two inelastic channels will effect only ${A}_{1/2}$, while the third will effect both ${A}_{1/2}$ and ${A}_{3/2}$. In  the following we enumerate the three decay amplitudes  ${B}^{+}\rightarrow {D}_{s}^{+} {\bar{D}}^{0}$,  ${B}^{+}\rightarrow \eta {K}^{+}$ and ${B}^{+}\rightarrow{\pi}^{0}{K}^{+}$.
\vskip 5mm
\begin{center}
{\bf ${B}^{+}\rightarrow \eta {K}^{+}$ Amplitude}
\end{center}

Due to $\eta - {\eta}^{\prime}$ mixing, the expression for the decay amplitude for  ${B}^{+}\rightarrow \eta {K}^{+}$ is fairly involved. In writing the following ${m}_{B}^{2} -
{m}_{K}^{2}\approx {m}_{B}^{2} - {m}_{\eta}^{2}
\approx {m}_{B}^{2}$ has been used ( a factor of ${{G}_{F} \over \sqrt{2}}$ is suppressed)
,
\begin{eqnarray}
A({B}^{+}\rightarrow {\eta} {K}^{+}) & = & {\xi}_{u}^{*} {m}_{B}^{2} \left[  {f}_{K}{F}_{0}^{B\eta}({m}_{K}^{2})F(\theta)  \left( {a}_{1} + {a}_{4}^{(u)}+ {R}_{u}^{(K)}{a}_{6}^{(u)}\right) \right. \nonumber \\
& & + {F}_{0}^{BK}({m}_{\eta}^{2}){f}_{\eta} \left\{ F(\theta){a}_{2} +{\mbox {sin}\theta \over \sqrt{3}} \left({a}_{5}^{(u)}  -{a}_{3}^{(u)}\right)\right.\nonumber \\ 
& & +G(\theta) \left.\left.\left({a}_{4}^{(u)} +{R}_{s}^{(\eta)}{a}_{6}^{(u)}\right)\right\}\right] \nonumber \\
& & + {\xi}_{c }^{*} {{m}_{B}}^{2}\left[ {f}_{K}{F}_{0}^{B\eta}({m}_{K}^{2})F(\theta)\left({a}_{4}^{(c)} +{R}_{u}^{(K)}{a}_{6}^{(c)}\right)\right.  \nonumber \\
& & +{F}_{0}^{BK} ({m}_{\eta}^{2}){f}_{\eta}\left\{{\mbox {sin}\theta \over \sqrt{3}}\left ({a}_{5}^{(u)}  -{a}_{3}^{(u)}\right)\right. \nonumber \\ & &  \left.\left.+G(\theta) \left({a}_{4}^{(u)} +{R}_{s}^{(\eta)}{a}_{6}^{(u)}\right)\right\}\right] 
\end{eqnarray}

where 
\begin{eqnarray}
{R}_{s}^{(\eta)} ={2 {m}_{\eta}^{2} \over ({m}_{b} -{m}_{s})(2{m}_{s})},~~~F(\theta)=\left( {1 \over \sqrt{6} } \mbox {cos}\theta-{1 \over \sqrt{3} } \mbox {sin}\theta\right), \nonumber \\
G(\theta)= -\left( {2 \over \sqrt{6}} \mbox {cos}\theta +{1 \over \sqrt{3} } \mbox {sin}\theta \right), ~~~~~~~~\theta
=  \eta - {\eta}^{\prime} ~\mbox {mixing angle}.
\end{eqnarray}

The largest contribution to the decay amplitude arises from the tree amplitudes proportional to ${a}_{1}$ and ${a}_{2}$ which constitute the coefficient of ${\xi}_{u}^{*}$. Numerically, with $\theta = -20^{\circ}$, and form factors from \cite{bsw},
\begin{eqnarray}
A({B}^{+}\rightarrow \eta {K}^{+})  & \equiv &  {\xi}_{u}^{*} {y}_{u}+  {\xi}_{c}^{*} {y}_{c} \nonumber \\ & = & {\xi}_{u}^{*} (0.8901 -0.00318 i) \nonumber \\
&  & + {\xi}_{c}^{*} (-0.640 - 0.277 i) 10^{-2}~GeV^3.
\end{eqnarray}
\vskip 5mm
\begin{center}
{\bf  ${B}^{+}\rightarrow {D}_{s}^{+} {\bar{D}}^{0}$ Amplitude}
\end{center}
In the factorization approximation one obtains ( a factor of ${{G}_{F} \over \sqrt{2}}$ is suppressed)

\begin{eqnarray}
A({B}^{+}\rightarrow {D}_{s}^{+} {\bar{D}}^{0}) & = &  ({m}_{B}^{2}-{m}_{D}^{2}) 
{f}_{{D}_{s}}
{F}_{0}^{BD}({m}_{{D}_{s}}^{2}) \left\{ {\xi}_{u}^{*} \left({a}_{4}^{(u)} + {R}_{c}^{({D}_{s})}{a}_{6}^{(u)}\right)\right. \nonumber \\
& & \left. + {\xi}_{c}^{*} \left({a}_{1}+{a}_{4}^{(c)}+{R}_{c}^{({D}_{s})}{a}_{6}^{(c)}\right)\right\}
\end{eqnarray}

where
\begin{equation}
{R}_{c}^{({D}_{s})} = {2 {m}_{{D}_{s}}^{2}\over {({m}_{b}-{m}_{c})({m}_{s}+{m}_{c})}}
\end{equation}

Numerically,
\begin{eqnarray}
A({B}^{+}\rightarrow {D}_{s}^{+} {\bar{D}}^{0}) & \equiv & {\xi}_{u}^{*}{z}_{u} + {\xi}_{c}^{*} {z}_{c} \nonumber \\ & =& {\xi}_{u}^{*} (-0.4758 - 0.2067 i) \nonumber \\
&  &+ {\xi}_{c}^{*} (4.5055 -0.1800 i){GeV}^{3}
\end{eqnarray}

We have used ${f}_{{D}_{s}}=0.3$ $GeV$ and ${{F}_{0}}^{BD}({{m}_{D}}^{2})$ =0.66 from Heavy Quark Effective Theory \cite{n1}.
\vskip 5mm
\begin{center}
{\bf ${B}^{+}\rightarrow{\pi}^{0}{K}^{+}$ Amplitude}
\end{center}
From Eq. (9),
\begin{equation}
A({B}^{+}\rightarrow{\pi}^{0}{K}^{+})=\sqrt{2}{A}_{3/2} -{1 \over \sqrt{2}}{A}_{1/2}^{(+)},
\end{equation}
where ${A}_{3/2}$ and ${A}_{1/2}^{(+)}$ have been defined in Eq. (15). 

\vskip 1cm
\begin{center}
{\bf IV. INELASTIC FINAL-STATE INTERACTIONS}
\end{center}
\vskip 5mm
\begin{center}
{\bf A. Formalism}
\end{center}

For the purposes of the following discussion, label channels ${\pi}^{+} {K}^{0}$, ${D}_{s}^{+} {\bar{D}}^{0}$,  $\eta {K}^{+}$ and ${\pi}^{0} {K}^{+}$ as channels 1, 2, 3 and 4, respectively. Let us also denote by the matrix ${S}_{ij}^{(I)}$ $(i,j=1,2,3,4)$ the elements of the $S$-wave scattering matrix in isospin $I$ state. Then the effect of rescattering on the weak decay amplitudes ${A}_{i}$ is to generate the unitarized amplitudes ${A}_{i}^{U}$ via \cite{kl}
\begin{equation}
{A}_{i}^{U(I)} = {\left({1+{S}^{(I)} \over 2}\right)}_{ij} {A}_{j}^{(I)},
\end{equation}
where ${A}_{i}^{(I)}$ are the weak decay amplitudes calculated so far in this paper according to the isospins. For the two isospin amplitudes, the above equation reads explicitly (note that the isospin label is now a subscript in the decay amplitudes but a superscript in the $S$ matrix elements),
\begin{eqnarray}
{A}_{3/2}^{U} & = &{1 \over 2} \left(1+{S}_{11}^{(3/2)}+\sqrt{2}{S}_{14}^{(3/2)} \right) {A}_{3/2}({B}^{+}\rightarrow {\pi}^{+} {K}^{0}), \nonumber \\
{A}_{1/2}^{U} & = &{1 \over 2} \left(1+{S}_{11}^{(1/2)}-{1 \over \sqrt{2}}{S}_{14}^{(1/2)} \right) {A}_{1/2}({B}^{+}\rightarrow {\pi}^{+} {K}^{0}) \nonumber \\
& & +\left({{S}_{12}^{(1/2)} \over 2}\right) {A}({B}^{+}\rightarrow {D}_{s}^{+} {\bar{D}}^{0}) \nonumber \\
& & +\left({{S}_{13}^{(1/2)}\over 2}\right) {A}^{(0)}({B}^{+}\rightarrow {\eta} {K}^{+}),
\end{eqnarray}
where we have used Eq. (22) in writing the contribution of channel 4.
The amplitudes  ${A}_{1/2}({B}^{+}\rightarrow {\pi}^{+} {K}^{0})$  and  ${A}_{3/2}({B}^{+}\rightarrow {\pi}^{+} {K}^{0})$ are given in Eq. (15), and   $A({B}^{+}\rightarrow {\eta} {K}^{+})$ and $A({B}^{+}\rightarrow {D}_{s}^{+} {\bar{D}}^{0})$ in Eqns. (18) and (21), respectively. The effect of the charge-exchange channel ${\pi}^{0} {K}^{+}$ has been simply to affect the following replacements:
\begin{eqnarray}
{S}_{11}^{(1/2)} ~~~~ & \rightarrow & ~~~~ {S}_{11}^{(1/2)} -{1 \over \sqrt{2} }{S}_{14}^{(1/2)}, \nonumber\\
{S}_{11}^{(3/2)} ~~~~ & \rightarrow & ~~~~ {S}_{11}^{(3/2)} +\sqrt{2} {S}_{14}^{(3/2)}.
\end{eqnarray}

	Our next task is to calculate the scattering-matrix elements. For this we use the Regge-exchange model including the Pomeron and the highest lying trajectories for each reaction. The $S$ matrix for the scattering from channel i to channel j in $L=0$ state is given by \cite{kl},
\begin{equation}
{S}_{ij} = {\delta}_{ij} +{i \over 8\pi \sqrt{{\lambda}_{i}{\lambda}_{j} } } \int_{{t}_{min}}^{{t}_{max}}{T(s,t) dt},
\end{equation}
where ${t}_{min}$ and ${t}_{max}$ are the minimum and maximum values of the invariant momentum transfer $t$, and ${\lambda}_{i}$ and ${\lambda}_{j}$ are the usual triangular functions: ${\lambda}(x,y,z)={(x^2+y^2+z^2 -2xy -2xz -2yz)}^{1/2}$ for channels i and j, respectively. For u channel exchanges, replace $t$ by $u$. In the following we consider the evaluation of ${S}_{ij}$.
\vskip 3mm
\begin{center}
{\bf B. Evaluation of ${S}_{11}$}
\end{center}
For the elastic scattering, ${\pi}^{+} {K}^{0} \rightarrow {\pi}^{+} {K}^{0}$, we consider only the Pomeron, the $\rho$ and $f$ trajectories. ${K}^{*}$ exchange in the u channel is not permitted in ${\pi}^{+} {K}^{0} \rightarrow {\pi}^{+} {K}^{0}$ scattering as the exchanged object is an exotic
$(q\bar{q}q\bar{q})$ state. A $(q\bar{q})$ exchange is, however, allowed in ${\pi}^{+} {\bar{K}}^{0} \rightarrow 
{\pi}^{+} {\bar{K}}^{0}$ scattering.

\begin{center}
{\bf Pomeron Exchange}
\end{center}
The elastic scattering  ${\pi}^{+} {K}^{0} \rightarrow {\pi}^{+} {K}^{0}$ via Pomeron exchange is parametrized by \cite{don}
\begin{equation}
{T}_{P}(s,t) = \beta(t) {\left({s \over {s}_{0}}\right)}^{{\alpha}_{P}(t)}{e}^{i\pi{\alpha}_{P}(t)/2},
\end{equation}
with \cite{don}
\begin{equation}
{\alpha}_{P}(t) = 1.08 + 0.25 t,
\end{equation}
and \cite{nard}
\begin{equation}
\beta(t) = \beta(0) {e}^{2.8t},
\end{equation}
where the invariant momentum transfer $t$ is expressed in units of $GeV^2$ and its coefficient in units of ${GeV}^{-2}$. The energy scale, $s_0$ is taken to be $1$  $GeV^2$ for  ${\pi}^{+} {K}^{0} \rightarrow {\pi}^{+} {K}^{0}$ scattering. The coupling constant $\beta(0)$ is fixed by the additive quark model, yielding $\beta(0) = 2{\beta}_{uu}+ 2{\beta}_{us}$, where the flavor subindices refer to the quark flavors between which the Pomeron is exchanged. Using ${\beta}_{us} = {2 \over 3}{\beta}_{uu}$ \cite{zheng} suggested by high energy $\pi p$ and $K p$  elastic scattering data, and ${\beta}_{uu} \approx 6.5$, we obtain $ \beta(0) \approx 22$. With these parameters we calculate,
\begin{equation}
 {i \over 8\pi{\lambda}_{11}}\int{{T}_{P}(s,t)  dt} = -0.313.
\end{equation}

\begin{center}
{\bf $\rho$ Exchange}
\end{center}

The $\rho$ exchange amplitude for  ${\pi}_{\alpha} {K} \rightarrow {\pi}_{\beta} {K}$ scattering is parametrized as
\begin{equation}
{T}_{\rho}^{(\alpha\beta)}(s,t) = {1 \over 2}[{\tau}_{\beta},{\tau}_{\alpha}] {T}_{\rho}(s,t),
\end{equation}
where ${\alpha}$ and ${\beta}$ are the charge indices of the initial and final state pions, respectively, and 
\begin{equation}
{T}_{\rho}(s,t) = {\beta}_{\rho}(0) {\left(1-{e}^{-i\pi{\alpha}_{\rho}(t)}\right) \over \Gamma({\alpha}_{\rho}(t)) \mbox {sin}\pi{\alpha}_{\rho}(t)} \left({{s \over {s}_{0}}}\right)^{{\alpha}_{\rho}(t)},
\end{equation}
where ${s}_{0}$ is again chosen to be $1$ $GeV^2$, and the $\rho$ trajectory is taken to be,
\begin{equation}
{\alpha}_{\rho}(t) = {\alpha}_{\rho}(0) + {\alpha}_{\rho}^{\prime}(t),
\end{equation}
with ${\alpha}_{\rho}(0) = 0.5$ and ${\alpha}_{\rho}^{\prime} = 0.848$ ${GeV}^{-2}$.

The isospin projection operators in $I=1/2$ and $I=3/2$ states are,
\begin{equation}
{Q}_{\alpha\beta}^{(1/2)} = {1 \over 3}{\tau}_{\beta} {\tau}_{\alpha}, ~~~~~{Q}_{\alpha\beta}^{(3/2)} = {\delta}_{\alpha\beta} - {1 \over 3}{\tau}_{\beta} {\tau}_{\alpha}.
\end{equation}

Thus,
\begin{equation}
 {1 \over 2}[{\tau}_{\beta},{\tau}_{\alpha}] =2 {Q}_{\alpha\beta}^{(1/2)} - {Q}_{\alpha\beta}^{(3/2)},
\end{equation}
and $\rho$ exchange contributes to the two isospin states $I=1/2$ and $I=3/2$ in the ratio $2:-1$.
The coupling constant ${\beta}_{\rho}(0)$ is determined by taking the limit $(s\rightarrow \infty, t\rightarrow {m}_{\rho}^{2})$ in Eqns. (31) and (32) for the reaction ${\pi}^{+} {K}^{0} \rightarrow {\pi}^{+} {K}^{0}$ and equating it to the perturbative graph contribution with the $\rho$ meson in the t channel. This leads to 
\begin{equation}
{\beta}_{\rho}(0) = {\pi {{\alpha}_{\rho}}^{\prime} \over 4}{g}_{VPP}^{2} = 25.1,
\end{equation}
where ${g}_{VPP}^{2}/{4\pi} = 3.0$, determined from the $\rho$ width, has been used. With these parameters and approximating $\Gamma({\alpha}_{\rho}(t)) \mbox {sin}\pi{\alpha}_{\rho}(t)\approx \Gamma({\alpha}_{\rho}(0)) \mbox {sin}\pi{\alpha}_{\rho}(0) =\sqrt{\pi}$,  one gets,
\begin{equation}
 {i \over 8\pi{\lambda}_{1}}\int{{T}_{\rho}(s,t) dt} = -0.0202 + 0.0191 i.
\end{equation}

\begin{center}
{\bf $f$ Exchange}
\end{center}

The $f$-exchange amplitude for   ${\pi}_{\alpha} {K} \rightarrow {\pi}_{\beta} {K}$ scattering is given by
\begin{equation}
{T}_{f}^{(\alpha\beta)}(s,t) = {\delta}_{\alpha\beta} {T}_{f}(s,t),
\end{equation}
where,
\begin{equation}
{T}_{f}(s,t) = {\beta}_{f}(0) {\left(1+{e}^{-i\pi{\alpha}_{f}(t)}\right) \over \Gamma({\alpha}_{f}(t))   \mbox {sin}\pi{\alpha}_{f}(t)} \left({{s \over {s}_{0}}}\right)^{{\alpha}_{f}(t)}.
\end{equation}
The scale factor ${s}_{0}$ is chosen to be $1$ $GeV^2$. Since ${\delta}_{\alpha\beta} = {Q}_{\alpha\beta}^{(1/2)} + {Q}_{\alpha\beta}^{(3/2)}$, $f$ exchange contributes in equal measure to both isospin states. We consider the $f$ trajectory to be parallel to the $\rho$ trajectory, ${\alpha}_{f}^{\prime}={\alpha}_{\rho}^{\prime}=0.848 ~{GeV}^{-2}$ and determine ${\alpha}_{f}(0)$ from the central value of $f$ meson mass, $1.275 ~ GeV$. This yields ${\alpha}_{f}(0)=0.6215$. We also considered a $f$ trajectory degenerate with the $\rho$ trajectory, but the effect of such a trajectory on the $CP$ asymmetry was not significantly different from that of non-dgenerate trajectory. Consequently, we present here only the discussion pertaining to a non-degenerate $f$ trajectory, albeit parallel to the $\rho$ trajectory. Further, we approximate $\Gamma\left({\alpha}_{f}(t)\right)\mbox {sin}\pi{\alpha}_{f}(t)\approx \Gamma\left({\alpha}_{f}(0)\right)\mbox {sin}\pi{\alpha}_{f}(0) = 1.3381$. The constant ${\beta}_{f}(0)$ is determined by taking the limit $(s\rightarrow \infty, t\rightarrow {{m}_{f}}^{2})$ in Eq. (38) (and (39)) and equating it to the perturbative expression for the  ${\pi}^{+} {K}^{0} \rightarrow {\pi}^{+} {K}^{0}$ scattering through $f$ exchange. The coupling of $f$ to pion pair and $K\bar{K}$ is determined from the branching ratios $B(f\rightarrow \pi\pi)$ and $B(f\rightarrow K\bar{K})$ given in \cite{pdg}. This results in ${\beta}_{f}(0)=-77.21$. With these parameters, we get
\begin{equation}
 {i \over 8\pi{\lambda}_{1}}\int{{T}_{f}(s,t) dt} = -0.1577-0.2952i.
\end{equation}

Using crossing matrices, the element ${S}_{11}$ in $I=1/2$ and $I=3/2$ states is given by
\begin{eqnarray}
{S}_{11}^{(1/2)} & = & 1+ {i \over 8\pi{\lambda}_{1}}\int{\left\{{T}_{P}(s,t) +{T}_{f}(s,t) +2{T}_{\rho}(s,t)\right\} dt}\nonumber \\
{S}_{11}^{(3/2)} & = & 1+{i \over 8\pi{\lambda}_{1}}\int{\left\{{T}_{P}(s,t) +{T}_{f}(s,t) -{T}_{\rho}(s,t)\right\} dt},
\end{eqnarray}
where the subindices refer to the exchanged trajectory.

	Putting the contributions of the Pomeron, the $\rho$ and the $f$ from Eqns. (30), (37) and (40) in Eq. (41), we obtain
\begin{equation}
 {S}_{11}^{(1/2)} =  0.4889-0.2570i,~~~~{S}_{11}^{(3/2)}= 0.5495 -0.3143i. 
\end{equation}
\vskip 3mm
\begin{center}
{\bf C. Evaluation of ${S}_{12}$}
\end{center}

The inelastic process ${\pi}^{+}{K}^{0}\rightarrow \eta {K}^{+}$, a pure $I=1/2$ reaction, proceeds via the exchange of ${K}^{*}$ in the u channel. The Regge amplitude for this process is given by 
\begin{equation}
{T}_{{K}^{*}}(s,u) = {\beta}_{{K}^{*}}(0) {\left(1-{e}^{-i\pi{\alpha}_{{K}^{*}}(u)}\right) \over \Gamma({\alpha}_{{K}^{*}}(u)) \mbox {sin}\pi{\alpha}_{{K}^{*}}(u)} \left({{s \over {s}_{0}}}\right)^{{\alpha}_{{K}^{*}}(u)}.
\end{equation}
The coupling constant  ${\beta}_{{K}^{*}}(0)$ is determined by taking the limit  $(s\rightarrow \infty, u\rightarrow {{m}_{K}^{*}}^{2})$ in Eq. (43) and equating it to the perturbative amplitude for   ${\pi}^{+}{K}^{0}\rightarrow \eta {K}^{+}$ process through ${K}^{*}$ exchange. Relating ${K}^{*}-K-\eta$ and ${K}^{*}-K-\pi$ couplings to the coupling ${g}_{VPP}$ through $SU(3)$ symmetry, we get
\begin{equation}
 {\beta}_{{K}^{*}}(0) ={\sqrt{3}  \over 4\sqrt{2} }{g}_{VPP}^{2}\pi {{\alpha}_{{K}^{*}}}^{\prime}
{s}_{0}\mbox {cos}{\theta},
\end{equation}
where $\theta$, the $\eta-{\eta}^{\prime}$ mixing angle, is taken to be $-20^\circ$. The scale parameter $s_0$ is chosen to be $1~GeV^2$ and the $K^*$ trajectory is chosen to be parallel to the $\rho$ trajectory (${\alpha}_{{K}^{*}}^{\prime}=0.848~{GeV}^{-2}$). This requires ${\alpha}_{{K}^{*}}(0) = 0.325$ for the ${K}^{*}$ mass to be $895~MeV$. With these parameters we obtain ${\beta}_{{K}^{*}}(0) = 57.78$. In projecting the $L=0$ partial wave according to Eq. (26), we use the approximation,
\begin{equation}
\Gamma\left({\alpha}_{{K}^{*}}(t)\right)\mbox {sin}\pi{\alpha}_{{K}^{*}}(t)\approx \Gamma\left({\alpha}_{{K}^{*}}(0)\right)\mbox {sin}\pi{\alpha}_{{K}^{*}}(0) = 2.345.
\end{equation}

We finally obtain
\begin{eqnarray}
{S}_{12}^{(1/2)} &= & {i \over 8\pi\sqrt{{\lambda}_{1}{\lambda}_{2}} }\int_{{u}_{min}}^{{u}_{max}}
{{T}_{{K}^{*}}(s,u) du} \nonumber \\
& = & (-0.70448 + 1.10972 i) {10}^{-2}.
\end{eqnarray}
\vskip 3mm
\begin{center}
{\bf D. Evaluation of ${S}_{13}$}
\end{center}

The reaction ${\pi}^{+} {K}^{0}\rightarrow {{D}_{s}}^{+} {\bar{D}}^{0}$ is a pure $I=1/2$ process. It proceeds through the exchange of $D^*$ in the t channel. (Here t channel is defined by the reaction ${\pi}^{+} D^0 \rightarrow {\bar{K}}^{0} {{D}_{s}}^{+}$; note also that ${{D}_{s}}^{*}$ cannot be exchanged in the u channel, ${\pi}^{+} {{D}_{s}}^{-} \rightarrow {\bar{K}}^{0}{\bar{D}}^{0}$ since the exchanged object has to be neutral.) The Regge amplitude is
\begin{equation}
{T}_{13}(s,t) =  {\beta}_{{D}^{*}}(0) {\left(1-{e}^{-i\pi{\alpha}_{{D}^{*}}(t)}\right) \over \Gamma({\alpha}_{{D}^{*}}(t)) \mbox {sin}\pi{\alpha}_{{D}^{*}}(t)} \left({{s \over {s}_{0}}}\right)^{{\alpha}_{{D}^{*}}(t)}.
\end{equation}
Now, the reaction  ${\pi}^{+} {K}^{0}\rightarrow {{D}_{s}}^{+} {\bar{D}}^{0}$ has a threshold close to $4~GeV$. Thus, we anticipate the scale factor $s_0$ to be much larger than $1~GeV^2$, as has been assumed so far. Further, if one puts $D^*$ ($J^P=1^+$) and ${{D}_{2}}^{*}$ ($J^P=2^+$) on the same linear trajectory, one gets a much flatter trajectory, ${\alpha}_{{D}^{*}}^{\prime}\approx 0.5~ GeV^{-2}$, than the $\rho$ trajectory \cite{{kl},{volk}}.
With this slope one finds ${\alpha}_{{D}^{*}}(0)=-1.02$. In contrast, had we chosen the ${D}^{*}$ trajectory to be parallel to the $\rho$ trajectory, we would have obtained ${\alpha}_{{D}^{*}}(0) = -2.426$, a much lower intercept and, consequently, a much smaller contribution to ${T}_{13}(s,t)$. In addition, we use the approximation
\begin{equation}
\Gamma({\alpha}_{{D}^{*}}(t)) \mbox {sin}\pi{\alpha}_{{D}^{*}}(t) \approx \Gamma({\alpha}_{{D}^{*}}(0)) \mbox {sin}\pi{\alpha}_{{D}^{*}}(0)=\pi,
\end{equation}
which is due to the cancellation of the pole of the Gamma function at argument $-1$ with the zero of the sine function at ${\alpha}_{{D}^{*}}(0) = -1$. The constant  ${\beta}_{{D}^{*}}(0)$ is obtained by taking the limit $(s\rightarrow \infty, t\rightarrow {m}_{{D}^{*}}^{2})$ in Eq. (47) and equating it to the perturbative graph with ${D}^{*}$ exchange. The coupling $({D}^{*}D\pi)$ is related to ${g}_{VPP}$ through $SU(4)$ symmetry. This results in $ {\beta}_{{D}^{*}}(0) = 118.44$. With these parameters, we obtain
\begin{eqnarray}
{S}_{13}^{(1/2)} & = & {i \over 8\pi\sqrt{{\lambda}_{1}{\lambda}_{3}} }\int_{{t}_{min}}^{{t}_{max}}
{{T}_{{D}^{*}}(s,t) dt} \nonumber \\
& = & (-0.22834 + 0.36992 i) {10}^{-2}.
\end{eqnarray}
\vskip 3mm
\begin{center}
{\bf E. Evaluation of ${S}_{14}$}
\end{center}
 Consider now the charge-exchange (CEX) reaction ${\pi}_{\alpha} {K}^{0}\rightarrow {\pi}_{\beta}{K}^{+}$. It proceeds through the exchange of the $\rho$ trajectory in the t channel and that of the ${K}^{*}$ trajectory in the u channel. Let us write these two contributions in the following form:
\begin{eqnarray}
\mbox{$\rho$ trajectory:} & ~~~~~~ & {1 \over 2}[{\tau}_{\beta},{\tau}_{\alpha}] {T}_{CEX}^{(\rho)}(s,t), \nonumber\\
\mbox{${K}^{*}$ trajectory:} & ~~~~~~ & {\tau}_{\alpha}{\tau}_{\beta}{T}_{CEX}^{({K}^{*})}(s,u),
\end{eqnarray}
where
\begin{eqnarray}
{T}_{CEX}^{(\rho)}(s,t) = {\beta}_{CEX}^{(\rho)}(0) {\left(1-{e}^{-i\pi{\alpha}_{\rho}(t)}\right) \over \Gamma({\alpha}_{\rho}(t)) \mbox {sin}\pi{\alpha}_{\rho}(t)} \left({{s \over {s}_{0}}}\right)^{{\alpha}_{\rho}(t)},\nonumber\\
{T}_{CEX}^{({K}^{*})}(s,u)=  {\beta}_{CEX}^{({K}^{*})}(0) {\left(1-{e}^{-i\pi{\alpha}_{{K}^{*}}(u)}\right) \over \Gamma({\alpha}_{{K}^{*}}(u)) \mbox {sin}\pi{\alpha}_{{K}^{*}}(u)} \left({{s \over {s}_{0}}}\right)^{{\alpha}_{{K}^{*}}(u)}.
\end{eqnarray}
The constants ${\beta}_{CEX}^{(\rho)}(0)$ and ${\beta}_{CEX}^{({K}^{*})}(0)$ are determined as in other cases heretofore, to be
\begin{equation}
 {\beta}_{CEX}^{(\rho)}(0) =2 {\beta}_{\rho}(0) = 50.21, ~~~~~~ {\beta}_{CEX}^{({K}^{*})}(0) = -{\beta}_{\rho}(0) = -25.1.
\end{equation}
Using the crossing matrices for the isospin spin operators, we obtain
\begin{eqnarray}
{S}_{14}^{(1/2)} & = & {i \over {8\pi{\lambda}_{1}}}\left(2\int_{{t}_{min}}^{{t}_{max}}{ {T}_{CEX}^{(\rho)}(s,t)dt} - \int_{{u}_{min}}^{{u}_{max}}{{T}_{CEX}^{({K}^{*})}(s,u)du}\right) \nonumber\\ & = & -0.08386+0.081207i 
\end{eqnarray}
\begin{eqnarray}
{S}_{14}^{(3/2)} & = & {i \over {8\pi{\lambda}_{1}}}\left(-\int_{{t}_{min}}^{{t}_{max}}{ {T}_{CEX}^{(\rho)}(s,t)dt }+ 2 \int_{{u}_{min}}^{{u}_{max}}{{T}_{CEX}^{({K}^{*})}(s,u)du}\right) \nonumber\\ & = & 0.046527-0.047815i. 
\end{eqnarray}

\vskip 1cm

\begin{center}
{\bf V. $CP$ ASYMMETRY CALCULATIONS}
\end{center}
\vskip 5mm
\begin{center}
{\bf A. Definitions and direct $CP$ asymmetry without long-range FSI}
\end{center}
\vskip 3mm
Let us define the generic form of the decay amplitude for  ${B}^{+}\rightarrow    {\pi}^{+} {K}^{0}$ as
\begin{equation}
A({B}^{+}\rightarrow {\pi}^{+} {K}^{0}) = {\xi}_{u}^{*} {A}_{u} +  {\xi}_{c}^{*} {A}_{c},
\end{equation}
then the decay amplitude for the process ${B}^{-}\rightarrow    {\pi}^{-} {\bar{K}}^{0}$ is
\begin{equation}
A({B}^{-}\rightarrow {\pi}^{-} {\bar{K}}^{0}) = {\xi}_{u} {A}_{u} +  {\xi}_{c} {A}_{c}.
\end{equation}
We define the direct $CP$ asymmetry as,
\begin{equation}
{A}_{CP}^{dir} = {{\Gamma({B}^{+}\rightarrow    {\pi}^{+} {K}^{0}) - \Gamma({B}^{-}\rightarrow    {\pi}^{-} {\bar{K}}^{0})} \over {\Gamma({B}^{+}\rightarrow    {\pi}^{+} {K}^{0}) + \Gamma({B}^{-}\rightarrow    {\pi}^{-} {\bar{K}}^{0})}}.
\end{equation}
Further, if we define the $CKM$ products with their weak phases following Wolfenstein parametrization,
\begin{equation}
{\xi}_{u}^{*} = \mid{\xi}_{u}\mid {e}^{i{\gamma}}, ~~~ {\xi}_{c} = \mid{\xi}_{c}\mid ,
\end{equation}
and the decay amplitudes ${A}_{u}$ and ${A}_{c}$ with their strong phases,
 \begin{equation}
{A}_{u} = \mid{A}_{u}\mid {e}^{i{\delta}_{u}}, ~~~ {A}_{c} = \mid{A}_{c}\mid {e}^{i{\delta}_{c}},
\end{equation}
then
\begin{equation}
{A}_{CP}^{dir} \approx {-2 r \xi \mbox {sin} ({\delta}_{u}-{\delta}_{c})  \mbox {sin} \gamma \over{1+ 2r\xi \mbox {cos} ({\delta}_{u}-{\delta}_{c})  \mbox  {cos}\gamma}} ,
\end{equation}
where we have defined,
\begin{equation}
\xi\equiv{\mid {\xi}_{u} \mid \over \mid {\xi}_{c} \mid} ={\mid {V}_{ub}{V}_{us}^{*} \mid \over \mid {V}_{cb}{V}_{cs}^{*} \mid} ={\lambda} {\mid {V}_{ub} \mid \over \mid {V}_{cb} \mid }, ~~~ \mbox{and} ~~ r= {\mid{A}_{u}\mid\over \mid{A}_{c}\mid},
\end{equation}
where $\lambda =0.22$ is the usual Wolfenstein parameter representing ${V}_{us}$, and in the denominator of Eq. (60), we have neglected a term of order ${\mid\xi\mid}^{2}$. Note also from the expression for ${A}_{CP}^{dir}$, Eq. (60) (which corresponds to the definition used in Ref. \cite{buras1}), that for a given value of the angle $\gamma$ the $CP$ asymmetry can be raised either by raising $r$, or by raising $\mbox {sin} ({\delta}_{u}-{\delta}_{c})$, or both. This point is discussed in the next section in the context of our calculations.

In the absence of  long-range FSI, the decay amplitude for  ${B}^{+}\rightarrow    {\pi}^{+} {K}^{0}$ is given by ${A}_{+0}$ of Eq. (12). One can then calculate the direct $CP$ asymmetry from the numerical values for ${A}_{3/2}$ and ${A}_{1/2}$ given in Eq. (15) with ${A}_{+0}= {A}_{3/2}+{A}_{1/2}$. Recall that the superscript $(+)$ in ${A}_{1/2}^{(+)}$ has been dropped.

With $\lambda=0.22$ and $\xi=0.022$ (see Eqns. (8) and (61)) and the three choices for $\gamma = (50^\circ, 75^\circ, 90^\circ)$, the $CP$ asymmetries calculated with  ${q}^{2}= {{m}_{b}}^{2}/2$ are (and shown in Table 1) is,
\begin{equation}
{A}_{CP}^{dir}  =  (-0.29\%, -0.37\%, -0.39\%), ~\mbox {for}~ \gamma = (50^\circ, 75^\circ, 90^\circ).
\end{equation}
 The reason for the asymmetry being an order of magnitude smaller than $\mid\xi\mid \sim O({\lambda}^{2})$ is that sin$({\delta}_{u}-{\delta}_{c})$ is of order ${10}^{-1}$ (see Table 1). Note also that the asymmetries in absence of FSI are comparable to those calculated in \cite{kram}. The authors of Ref. \cite{kram} have used  a different set of mass parameters; the phase ${\delta}_{c}$ is sensitive to the choice of $m_c$, particularly when ${m}_{c}^{2}$ is in the vicinity of ${m}_{b}^{2}/8$. In fact, the $CP$ asymmetry is quite sensitive to the choice of $q^2$ (see Ref. \cite{kram}). The difference in the sign of the asymmetry between us and Ref. \cite{kram} is a matter of definition only.
\vskip 3mm
\begin{center}
{\bf B. $CP$ asymmetry with long-range inelastic FSI}
\end{center}

In terms of the two isospin amplitudes, ${A}_{3/2}$ and ${A}_{1/2}^{(+)}$, defined in Eq. (14), the decay amplitude for ${B}^{+}\rightarrow {\pi}^{+} {K}^{0}$ is given by \begin{equation}
{A}_{+0}\equiv A({B}^{+}\rightarrow {\pi}^{+} {K}^{0})= {A}_{3/2}+{A}_{1/2}^{(+)}.
\end{equation}
 After unitarization, the two relevant isospin amplitudes are given by  Eq. (24).  Substituting from Eq. (24) in Eq. (63) and reorganizing the resulting expression, we obtain the decay amplitude after the inclusion of long-range inelasic FSI as,
\begin{equation}
{A}_{+0}= {\xi}_{u}^{*}{A}_{u}^{U}+ {\xi}_{c}^{*}{A}_{c}^{U},
\end{equation}
where
\begin{eqnarray}
{A}_{u}^{U} & = & {1 \over 2}\left(1+{S}_{11}^{(3/2)}+\sqrt{2}{S}_{14}^{(3/2)} \right) {x}_{u}^{(3/2)}+ {1 \over 2}\left(1+{S}_{11}^{(1/2)} -{1 \over \sqrt{2} }{S}_{14}^{(1/2)}\right) {x}_{u}^{(1/2)}  \nonumber \\
& & +\left({{S}_{12}^{(1/2)} \over 2}\right) {y}_{u}
+\left({{S}_{13}^{(1/2)}\over 2}\right) {z}_{u} \nonumber \\
{A}_{c}^{U} & = &{1 \over 2} \left(1+{S}_{11}^{(1/2)}-{1 \over \sqrt{2} }{S}_{14}^{(1/2)}\right) {x}_{c}^{(1/2)}  \nonumber \\
& & +\left({{S}_{12}^{(1/2)} \over 2}\right) {y}_{c}
+\left({{S}_{13}^{(1/2)}\over 2}\right) {z}_{c}.
\end{eqnarray}

The weak decay parameters $({x}_{u}^{(1/2)}, {x}_{u}^{(3/2)}, {y}_{u}, {z}_{u})$ and $({x}_{c}^{(1/2)},{y}_{c},{z}_{c})$ are given in Eqns. (15), (18) and (21). The $S$-matrix elements ${S}_{11}^{(1/2)}, {S}_{11}^{(3/2)}, {S}_{12}$, ${S}_{13}$, ${S}_{14}^{(1/2)}$ and ${S}_{14}^{(3/2)}$ are given in Eqns. (42), (46),  (49), (53) and (54), respectively. The $CP$ asymmetry is now given by 
\begin{equation}
{A}_{CP}^{dir} \approx {-2 {r}^{\prime} \xi \mbox {sin} ({\delta}_{u}^{\prime}-{\delta}_{c}^{\prime}) \mbox {sin}\gamma \over{1+ 2{r}^{\prime}\xi \mbox {cos} ({\delta}_{u}^{\prime}-{\delta}_{c}^{\prime}) \mbox {cos}\gamma}},
\end{equation}

where
\begin{equation}
{A}_{u}^{U} = \mid{A}_{u}^{U}\mid {e}^{i{\delta}_{u}^{\prime}}, ~~~ {A}_{c}^{U} = \mid{A}_{c}^{U}\mid {e}^{i{\delta}_{c}^{\prime}}, ~~~ {r}^{\prime}= {\mid {A}_{u}^{U}\mid \over \mid{A}_{c}^{U}\mid}.
\end{equation}

The resulting $CP$ asymmetries calculated with ${q}^{2}={{m}_{b}}^{2}/2$ are:
\begin{equation}
{A}_{CP}^{dir}   =   (0.73\%,0.94\%,0.99\%), ~\mbox {for}~ \gamma = (50^\circ, 75^\circ, 90^\circ)
\end{equation}

Comparison with the $CP$ asymmetries in absence of FSI shows that $CP$ asymmetries have increased approximately three-fold, and have changed signs.

We studied another scenario - one where channels 2 and 3 were removed and only channels 1 and 4 retained. In such a scenario only the elastic and the CEX channels come into play and the effect of the FSI are embodied in ${S}_{11}^{(I)}$ and ${S}_{14}^{(I)}$. The results were,
\begin{equation}
{A}_{CP}^{dir}   =  (0.50\%,0.65\%,0.68\%), ~\mbox {for}~ \gamma = (50^\circ, 75^\circ, 90^\circ)
\end{equation}
We note from Eqns. (68) and (69) that adding channels 2 and 3 has made a significant difference   to the asymmetry. 

	In yet another scenario in which only channel 1 was kept (only elastic re-scattering allowed), the $CP$ asymmetries were,
\begin{equation}
{A}_{CP}^{dir}   =   (0.44\%,0.56\%,0.59\%), ~\mbox {for}~ \gamma = (50^\circ, 75^\circ, 90^\circ).
\end{equation}

The results are summarized in Table 1.
 \vskip 1cm
\begin{center}
{\bf VI. DISCUSSION}
\end{center}
\vskip 5mm
In the absence of FSI, the decay ${B}^{+}\rightarrow {\pi}^{+} {K}^{0}$ is a pure penguin process; the decay amplitude depends on the coefficients $a_i (i=3, ... ,6)$. With our choice of the parameters, and with an effective ${q}^{2}={m_b}^{2}/2$, we have calculated the direct $CP$ asymmetry ${A}_{CP}^{dir}$ in the absence of FSI to be in the range $(-0.3 ~\mbox {to} -0.4)\%$, depending on the choice of  $\gamma$. This is an order of magnitude smaller than ${\lambda}^{2}$.

Once the decay amplitude for ${B}^{+}\rightarrow {\pi}^{+} {K}^{0}$ is decomposed into the two isospin states, $I=1/2$ and $I=3/2$, each isospin amplitude involves the tree coefficients $a_1$ and $a_2$; it is just that they are cancelled in the sum. As the rescattering $S$ matrices ${S}^{(1/2)}$ and ${S}^{(3/2)}$ are different, the cancellation of the coefficients $a_1$ and $a_2$ does not occur after FSI are taken into account. That is, FSI mix and blur the classification of processes by their topologies, tree or penguin.

Considerable attention has been paid \cite{{jin},{du},{falk}} to the size of the direct $CP$ asymmetry in  ${B}^{+}\rightarrow {\pi}^{+} {K}^{0}$. It has been argued in \cite{falk} that asymmetries as large as $20\%$ could be generated as a result of FSI. Refs. \cite{{jin},{du}} argue that $10\%$ asymmetry could be produced in the mechanism they propose. Our calculations suggest otherwise. We expand on a critique of these papers, and a related one \cite{delepine}, in the following. 

In order to discuss the results of \cite{falk}, we need to establish the correspondence between their notation and ours:
\begin{eqnarray}
\mbox {Before FSI:} ~~~~~~~~~~
 \mid {r}_{+} \mid ~~(\mbox{Ref.} \cite{falk}) & = & \mid r\xi \mid ~~ (\mbox {here} \nonumber \\
& = & {\mid {V}_{ub}{{V}_{us}}^{*} \mid \over \mid {V}_{cb}{{V}_{cs}}^{*} \mid} {\mid {A}_{u} \mid \over \mid {A}_{c} \mid} \nonumber \\
& = & \lambda {\mid {V}_{ub} \mid \over \mid {V}_{cb} \mid}
{\mid {A}_{u} \mid \over\mid {A}_{c}\mid}.
\end{eqnarray}

Consequently, since $\mid {A}_{u} \mid /\mid {A}_{c} \mid \approx 1$, see Table 1, and $\mid {V}_{ub} \mid /\mid {V}_{cb} \mid = 0.08\pm 0.02$ \cite{pdg}, one finds $\mid {r}_{+} \mid <\lambda/10$. After FSI are taken into account, the correspondence reads,
\begin{equation}
\mbox {After FSI:}~~~~~~~~~~ \mid \epsilon \mid ~~(\mbox{Ref.} \cite{falk}) =  \lambda {\mid {V}_{ub} \mid \over \mid {V}_{cb} \mid}
{\mid {{A}_{u}}^{\prime} \mid \over\mid {{A}_{c}}^{\prime}\mid}~~ (\mbox {here}).
\end{equation}

It is argued in \cite{falk} that $\epsilon$ could be as large as $0.09$ as a result of FSI. For this to be achieved one would need $\mid {{A}_{u}}^{\prime} \mid/\mid {{A}_{c}}^{\prime} \mid > 4$. Our calculations show (see Table 1) that after incorporating inelastic FSI, this ratio has remained close to unity. As a consequence, on this score alone, our calculated $CP$ asymmetry would be a factor of four lower.  A further lowering (by a factor of five) comes about as follows: Table 1  shows that the strong phase factor $\mbox {sin} ({\delta}_{u}-{\delta}_{c})$, achieves a value of  $\approx 0.2$ in magnitude. In contrast, this factor is not estimated in Ref. \cite{falk}, and an optimistic value of unity is assumed for this factor. This provides a further decrease in the $CP$ asymmetry by a factor of five. Thus, our calculated $CP$ asymmetry is a factor of 20 lower than that in Ref. \cite{falk}.

The estimates of \cite{falk}, which appear to be too optimistic, could be criticized on several grounds. First, to estimate the effect of FSI, they calculate the discontinuity of the decay amplitude across the unitarity cut using a Regge amplitude for the scattering amplitude. A dispersion relation is then used to generate the FSI-corrected amplitude. Their final result (Eqn. (2.17) of Ref. \cite{falk}) amounts to retaining only the absorptive part of the integral (pole contribution) and neglecting the dispersive part (coming from the principal part integration). Besides, the integration ranges from $s={({m}_{K}+{m}_{\pi})}^{2}$ to infinity; Regge description hardly applies at the lower end of the integration range. Secondly, having estimated the effect of ${\pi}^{0}{K}^{+}$ state through the exchange of the $\rho$ trajectory on the parameter $\epsilon$  (see Eq. (2.25) of \cite{falk}), the contributions of other channels are estimated only very crudely (see material following Eq. (2.25) of \cite{falk}). Our estimates do not support the optimistic expectations that inelastic channels add up coherently to generate an $\epsilon$ of the size required to produce a (10 - 20)\% $CP$ asymmetry.

An estimate of a $10\%$ $CP$-asymmetry in  ${B}^{+}\rightarrow {\pi}^{+} {K}^{0}$ in Refs. (\cite{jin},\cite{du}) is based on the calculation of the triangle graph representing the decay ${B}^{+}\rightarrow {\pi}^{+} {K}^{0}$ via an intermediate $\rho{K}^{*}$ state. The transition $\rho{K}^{*}\rightarrow {\pi}^{+}{K}^{0}$ is assumed to occur through an elementary $\pi$ exchange. This is tantamount to saying that one-pion-exchange mechanism  reasonably describes the reaction $\rho{K}^{*}\rightarrow {\pi}^{+}{K}^{0}$ at $\sqrt{s}\approx 5~GeV$. That this is not so is discussed below.

Consider the scattering of helicity zero ${\rho}^{+}$ and ${K}^{*0}$ into ${\pi}^{+} {K}^{0}$. The scattering $T$-matrix for this reaction with the exchange of a ${\pi}^{0}$ is given by
\begin{equation}
T(s,t) = { 2 {{g}_{VPP}^{2}}\over {m}_{\rho}{{m}^{*}}_{K} (t-{m}_{\pi}^{2})} (p{E}_{\pi}-k{E}_{\rho}\mbox {cos}\theta)(p{E}_{K}-k{E}_{{K}^{*}}\mbox {cos}\theta),
\end{equation}
where $t= {({p}_{\pi}-{p}_{\rho})}^{2} = {{m}_{\rho}^{2}}+{{m}_{\pi}^{2}}-2{E}_{\rho}{E}_{\pi}+2pk \mbox {cos}\theta$. Here $\theta$ is the scattering angle in the center of mass, and $k$ and $p$ are the center of mass momenta in the ${\pi}^{+}{K}^{0}$ and ${\rho}^{+}{K}^{*0}$ channels, respectively.

For zero-helicity scattering, the rotation matrices ${d}^{J}_{00}(\theta)$ are the same as the Legendre polynomials ${P}_{J}( \mbox {cos} {\theta})$. Hence the projection of $J=0$ partial wave proceeds as in Eq. (26). Converting the variables from the invariant momentum transfer $t$ to cos${\theta}$, one gets
\begin{eqnarray}
{S}_{J=0}({\rho}{K}^{*0} \rightarrow {\pi}^{+}{K}^{0}) & =  & i{\sqrt{pk}  \over 8\pi\sqrt{s} }
\int_{-1}^{+1}{T(s, \mbox {cos}{\theta}}) d \mbox {cos} {\theta} \nonumber \\
 &  &  \approx -28 i.
\end{eqnarray}
Unitarity requires that the off-diagonal (and the diagonal) elements of the $S$ matrix be bounded by unity. Thus, the one-pion-exchange mechanism leads to a gross violation of unitarity of the $S$ matrix. It is for this reason that a large $CP$ asymmetry is generated in \cite{jin} and \cite{du}.

Finally, we would like to comment on a related work of Ref. \cite{delepine}. In this paper, the authors evaluate the isospin $1/2$ and $3/2$ scattering amplitudes in $\pi K\rightarrow \pi K$; the mechanism is assumed to be through the exchanges of the Pomeron and Regge ($\rho$, ${K}^{*}$, etc.) trajectories. The partial wave amplitudes are then projected and the scattering phases, ${\delta}_{l}$ identified as (see following Eq. (26) of \cite{delepine}):
\begin{equation}
\mbox {tan}{\delta}_{l} = {Im {A}_{l}(s) \over Re {A}_{l}(s)}.
\end{equation}
This identification is correct only if the scattering is elastic, i.e. the $S$ matrix is given by ${e}^{2i{\delta}_{l}}$. To wit:
\begin{eqnarray}
\mbox {If} ~~~~~~~~~~~~S_l  & =  & e^{2i{\delta}_{l}} \nonumber \\
\mbox {then}~~~~~~~~ A_l  & \equiv & {1 \over 2i}(S_l - 1) \nonumber \\ & = & \mbox {sin}{\delta}_{l} e^{i{\delta}_{l}},
\end{eqnarray}
where $A_l$ is the scattering matrix constrained by unitarity of the $S$ matrix. Clearly now, Eq. (75) holds. However, if the scattering is inelastic, as we anticipate it to be in the $B$-mass region, then each diagonal element (ii) of the $S$ matrix can be written as,
\begin{equation}
{S}_{ii}^{l} = {\eta}_{i}e^{2i{\delta}_{i}^{l}},
\end{equation}
where the elasticity ${\eta}_{i}$ is such that $0 < {\eta}_{i} < 1$. Now,
\begin{eqnarray}
{A}_{i}^{l} & \equiv & {1 \over 2i}({S}_{ii}^{l} - 1) \nonumber \\
		& = & {1 \over 2i}({\eta}_{i} e^{2i{\eta}_{i}^{l}} - 1),
\end{eqnarray}
and one no longer obtains Eq. (75). Thus, we believe, the identification of the scattering phase shift in \cite{delepine} is incorrect. However, using the real and the imaginary parts of the calculated amplitude ${A}_{i}^{l}$, one can calculate both ${\eta}_{i}$ and ${\delta}_{i}^{l}$.

So much about the scattering amplitude and its phase, the phase of the weak decay amplitude is quite a different matter. Consider a spinless single channel (elastic) scattering, and a single weak decay amplitude, $A(s)$. The $S$ matrix in $L=0$ state is $S = e^{2i {\delta}}$. Then the unitarized weak decay amplitude ${A}^{U}(s)$ is,
\begin{eqnarray}
{A}^{U}(s) & = & \left({ 1 + S\over 2}\right) A(s) \nonumber \\
		&   =   & \mbox {cos}{\delta}(s) A(s) e^{i{\delta}(s)}.
\end{eqnarray}
Clearly, the phase of the weak decay amplitude is the scattering phase (Watson's theorem) if the decay amplitude is real. The claim of \cite{delepine} that the phase they are calculating is the phase of the weak-decay amplitude, therefore, is correct only if the scattering is elastic. The scattering, however, is not elastic as evidenced by $\mid {S}_{ii} \mid < 1$ in our calculation.

In general, for inelastic scattering, the unitarized weak decay amplitude is a superposition of the form given in Eq. (23), and the phase of ${A}_{i}^{U}$ is, obviously, not simply related to the phases of the $S$-matrix elements, but to the $S$-matrix elements weighted by the weak-decay amplitudes ${A}_{j}$.

A final word of caution: Our calculations have been done within the ambit of a few, hopefully important, two-body channels. However, our results contradict the conclusions of other calculations with the same limitations. Many-body decay channels could, in principle, play an important role in determining the $CP$ asymmetry; yet their effect is difficult to estimate. Thus, whether large $CP$ asymmetries could be generated within the Standard Model remains an open question. 
\vskip 5mm

 An Individual Operating grant from the Natural Sciences and Engineering Research Council of Canada  partially supported this research.
\newpage

\begin{table}
\begin{center}
\caption {The ratio $r= \mid {A}_{u} \mid / \mid{A}_{c} \mid $, the strong phase factor sin$({\delta}_{u}-{\delta}_{c})$, and $CP$ asymmetries in \% in different scenarios.}
\vskip 3mm

\begin{tabular}{|c|c|c|c|c|c|}
\hline\hline
Scenario  &  r  & sin$({\delta}_{u}-{\delta}_{c})$ & \multicolumn{3}{c|} {$CP$ asymmetry} \\ \cline{4-6}  
		&   =$\mid {A}_{u} \mid /\mid {A}_{c} \mid$  &  &  $\gamma=50^\circ$  & $\gamma=75^\circ$ &  $\gamma=90^\circ$ \\
\hline
No FSI   & 0.898 & 0.098 &  -0.29  & -0.37 & -0.39 \\
\hline
 Elastic FSI   & 1.071 & -0.125 & 0.44  & 0.56 & 0.59 \\
\hline
Elastic+CEX   & 1.088& -0.142  & 0.50  &  0.65 & 0.68  \\
\hline
All four channels & 1.103 & -0.203 & 0.73 & 0.94 & 0.99 \\
\hline\hline
\end{tabular}
\end{center}
\end{table}

\end{document}